# Nonvolatile Electrochemical Random-Access Memory Under Short Circuit

Diana Kim[1,2], Virgil Watkins[1], Laszlo Cline[1], Jingxian Li[1], Kai Sun[1],
Joshua D. Sugar[3], Elliot J. Fuller[3], A. Alec Talin[3], Yiyang Li[1#]

[1]Materials Science and Engineering, University of Michigan, Ann Arbor, MI, USA

[2]Macromolecular Science and Engineering, University of Michigan, Ann Arbor, MI, USA

[3] Sandia National Laboratories, Livermore, CA, USA

#Corresponding author: yiyangli@umich.edu

**Abstract**: Electrochemical random-access memory (ECRAM) is a recently developed and highly promising analog resistive memory element for in-memory computing. One longstanding challenge of ECRAM is attaining retention time beyond a few hours. This short retention has precluded ECRAM from being considered for inference classification in deep neural networks, which is likely the largest opportunity for in-memory computing. In this work, we develop an ECRAM cell with orders of magnitude longer retention than previously achieved, and which we anticipate to exceed 10 years at 85°C. We hypothesize that the origin of this exceptional retention is phase separation, which enables the formation of multiple effectively equilibrium resistance states. This work highlights the promises and opportunities to use phase separation to yield ECRAM cells with exceptionally long, and potentially permanent, retention times.

1. **Introduction**

In-memory computing using analog resistive nonvolatile memory is highly attractive for data-heavy workflows like artificial intelligence[1–4]. By co-locating the memory and processing elements on a single device, in-memory analog computing is potentially much faster and more energy efficient than optimized digital computers[1,5]. A grand challenge has been to identify the best analog resistive memory element at the core of in-memory computing; proposed technologies include floating gate memory[6], phase-change memory (PCM)[7–9], ferroelectric transistors[10,11], resistive redox memory (ReRAM)[12–17], and magnetic tunnel junction memory[18]. Unfortunately, none of these technologies have been able to meet all the demands of in-memory computing.

Electrochemical random-access memory, or ECRAM, is a promising recently developed device for analog in-memory computing[19–45]. ECRAM stores analog information states by electrochemically shuttling guest dopant ions like lithium ions, protons, or oxygen vacancies between two mixed-conducting host materials. Since research in ECRAM has dramatically increased from 2017[21,22], it has met many of the requirements for analog in-memory computing, including linear & symmetric among hundreds of analog states; low read and write currents, voltages, and energies; parallel weight updates[23,35]; CMOS-compatible materials with back-end-of-the-line (BEOL) thermal budgets[27,28,46]; and scaled sub-micron sized devices[24,28]. Recently, sub-microsecond switching times have been shown using protons-based into polymers[23,47], MXenes[34], and tungsten oxides[28]. These developments make ECRAM highly promising for on-line training, which aims to identify the correct analog resistance states (or weights) of a network using the training set of pre-classified data. An even larger application of in-memory computing is inference, which uses a pre-trained network to classify new data. Inference is not only simpler to conduct in analog, but is more ubiquitous because all distributed devices, including ones with very limited energy budgets, would use inference to classify new data.

An overarching challenge that has precluded ECRAM from inference accelerators is insufficient retention, generally only several hours at room temperature[22,25–27], and much lower than the 10 years at 85°C benchmark for ReRAM, PCM, and floating gate memory. Moreover, nearly all retention studies utilize an electronic switch that keeps the gate and channel electrodes under "open circuit;" while this switch can provide several hours of retention in larger devices >10 μm, leakage currents in the switch will preclude retention times in scaled sub-micron devices[25]. A recent study using a 100-nm proton-based ECRAM shows only about 100 s retention time due to this scaling-retention challenge[28]. We have previously attempted to solve this problem by replacing highly mobile protons and lithium ions with the immobile oxygen ion[25], effectively creating an "ionic switch" that can retain state under short circuit; however, this device would only retain state for about 1 day at 25°C, and degrades rapidly to only ~100 seconds at 70°C under short circuit. Due to these challenges, it is believed that ECRAM will not be able to meet the needs of analog inference.

In this work, we demonstrate a nonvolatile ECRAM with a short-circuit retention time several orders of magnitude higher than previously shown in ECRAM, and can likely meet the 10 years at 85°C benchmark for binary states. By analyzing the electrochemical currents, we believe that this unprecedented retention time arises from phase separation into oxygen-rich and oxygen-poor phases of the amorphous tungsten oxide materials used in the ECRAM cell. This yields nonvolatile change in the ion concentration that does not fully reverse upon applying short circuit conditions after switching. Although additional material selection and device engineering are needed to realize the requirements necessary for an inference engine, our work shows that harnessing phase separation can provide a powerful means to enable nonvolatile ECRAM to attain the necessary retention times for analog inference.

## 2. Results and Discussion

### 2.1 Tungsten suboxide ECRAM devices

Our ECRAM cells consist of 140 nm of amorphous tungsten sub-oxide ($WO_{3-X}$) grown on opposite sides of a single-crystal yttria-stabilized zirconia (YSZ) substrate using reactive sputtering (Fig. 1a). Supporting Fig. S1 uses the X-ray diffraction pattern to confirm the $WO_{3-X}$ material is amorphous. Three Pt current collectors are grown by sputtering over shadow masks (Supporting Fig. S2). While tungsten oxide can intercalate several ions, our use of the YSZ electrolyte means that this ECRAM cell operates on the basis of oxygen vacancy ions. Finally, a passivation layer of $HfO_X$ is grown to reduce the oxidation of $WO_{3-X}$ with the environment. Except for the YSZ, all processes are back end of the line (BEOL) compatible, with growth below 200°C. Energy dispersive spectroscopy (Supporting Fig. S3) of this sub-oxide suggests a metal to oxygen ratio of 2:5, yielding $WO_{3-X}$ where X~0.5. We operate this device by applying a positive (negative) gate voltage ($V_G$, Fig. 1a) to potentiate (depress) or negative gate voltage; during retention, we ground the gate by applying 0V, short-circuiting it to the channel. Because we do not use a selector or a switch with finite leakage in series with the gate, as done in most past research, our ability to retain state under short circuit can be scaled to smaller devices[25].

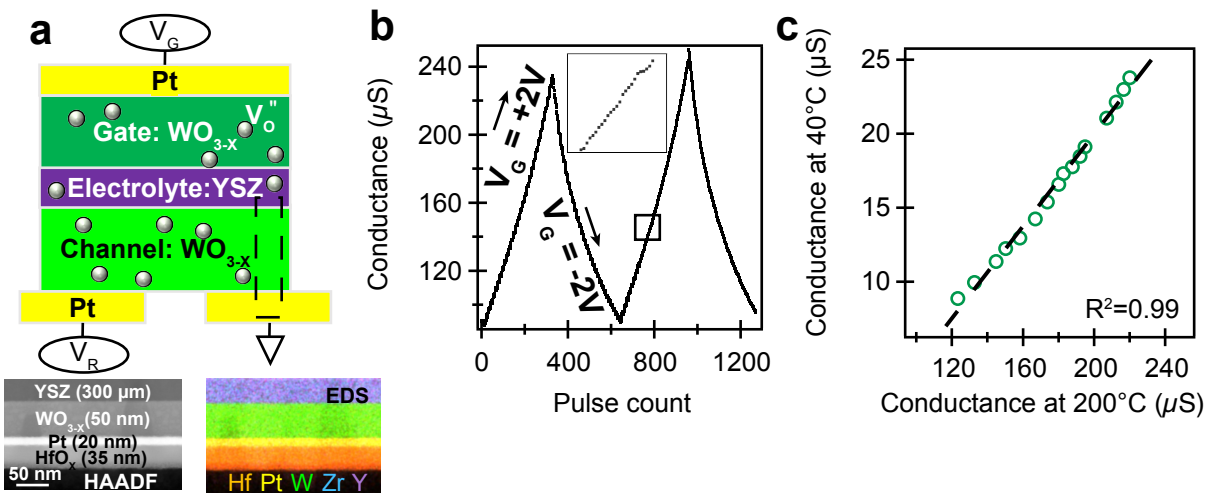

**Fig. 1: Tungsten suboxide based ECRAM cell used in this work.** (a) Schematic and cross-section TEM of the $WO_{3-X}$ ECRAM cell. This cell stores and switches analog information states as the concentration of oxygen vacancy ions in the channel. (b) Linear and symmetric switching among many analog states. Each pulse applies a voltage of +/- 2 V for 30 seconds. (c) There exists a direct linear mapping of the conductance at 200°C and the conductance at 40°C.

Like other ECRAM cells, this device possesses linear and symmetric analog switching, with at least 200 distinct analog states (Fig. 1b). The write current density of $10^{-14}$ A/μm$^2$ is comparable with other ECRAM cells[25], and much lower than that of phase-change, ReRAM, and floating gate memory. The channel conductance, at ~100 μS, is generally higher than optimal for in-memory computing[48]. However, the channel conductance drops by about an order of magnitude between 200°C to 40°C (Fig. 1c). As a result, while the conductance state should be set at high temperatures, inference could be conducted at near-room temperatures with much lower electronic conductances and consequently improved power efficiencies. Further decrease of the conductance by at least one order of magnitude can be achieved by simply reducing the width-to-length ratio from 16:1 to 1:1, and would place this cell well within the target of ~1 MΩ for analog inference. The switching time and temperature can be decreased by reducing the YSZ thickness from ~300 μm to <1000 nm, as has been shown in our past work[25].

## 2.2 Demonstration of Nonvolatile ECRAM

Having shown that we can switch to any resistance state within the range by applying a gate voltage for a given amount of time (Fig. 1b), we next investigate the retention behavior under short circuit. In Fig. 2a, we monitor the channel conductance when $V_G$=0V after the device is set to different resistance states. To our surprise, after a volatile transient response, the conductance states ultimately settle at different values, and not to a global equilibrium value as shown in other ECRAM[25,49]. A tangent line extrapolation over the last 300 seconds confirms that the channel conductances are not converging. While we only show five analog states due to experimental time constraints, we anticipate that each switched state in Fig. 1b yields a distinct retention profile and a different final conductance. Whereas applying a gate voltage of 0V without a switch to other ECRAM erases the previously programming resistance state (Supporting Fig. S4), this ECRAM cell shows that a non-volatile change in resistance can be achieved even under short circuit at elevated temperatures.

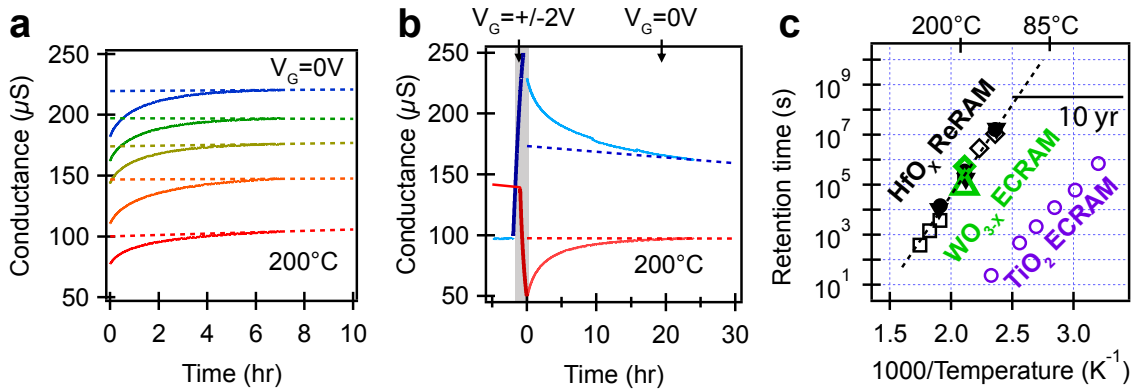

**Fig. 2: ECRAM retention characteristics when the gate and channel are shorted.** (a) After setting to a given state, our ECRAM cells will stabilize into one of several different analog resistance states. Dashed lines are tangent line constructions that project different conductance values as a function of time. (b) Retention times held for 24 hours from the low-resistance and high-resistance states. t = 0 hr signify when retention measurements start, which occur after a 1 or 2 hr switching at $V_G=\pm2V$. The tangents lines to these retention profiles converges after about 150 hours at 200°C. (c) Comparison of retention times of $WO_{3-X}$ ECRAM with filament-based ReRAM[50–53] and past $TiO_2$-based ECRAM[25]. Each different marker shape for $HfO_X$ ReRAM indicates results from a different paper. The two $WO_{3-X}$-ECRAM retention times at 24 (triangle) and 150 (diamond) hours equals the time for the experiment in (b) and the extrapolated tangent line convergence. Both values likely underestimate the retention time.

In Fig. 2b, we monitor non-volatility over 24 hours at 200°C from the high resistance and the low resistance state with $V_G=0V$. Once again, we observe a volatile transient response and a nonvolatile response. Interestingly, while the conductance of the high-resistance state is very stable, there is a slight decrease in the conductance of the low-resistance state over time. We hypothesize this is a result of slow atmospheric oxidation of the suboxides, despite the $HfO_2$ protection layer. However, even a tangent line extrapolation suggests that the high- and low-resistance states will not converge until at least 150 hours at 200°C. We note that the slopes of these tangent lines decrease over time and approach 0. If the tangent line reaches 0, the conductance values will not converge, and the projected retention approaches infinity. In the absence of atmospheric oxidation, we anticipate that the tangent line slope will eventually reach 0, and that the cell will exhibit a permanent change in the channel conductance. However, further work will be necessary to confirm the ultimate retention times, especially given the convoluting effects of oxidation.

We compare the retention time of our $WO_X$-ECRAM with that of other ECRAM cells and with two-terminal filamentary ReRAM based on $HfO_X$ (Fig. 2c). Our retention times are about 3-4 orders of magnitude higher than our previous anatase $TiO_2$-based ECRAM cells using a similar single-crystal YSZ electrolyte substrate[25]. More excitingly, both our achieved retention time of 24 hours and tangent-line extrapolated result of >150 hours at 200°C are comparable to that of $HfO_X$-based two-terminal ReRAM devices at the same temperature. It is well accepted that these ReRAM devices yield 10 years of retention time at 85°C[50–53].

We next aim to project the retention time to lower temperatures. In our device, the low-resistance and high-resistance states do not converge in 24 hours at 200°C, and may never converge. Thus, it is not possible to directly extrapolate the retention time by measuring when the high- and low-resistance states converge at different temperatures. Instead, we treat the gate electrochemical current $I_G$ as proportional to the ionic conductivity of the YSZ electrolyte, as shown in previous work[25]; in Supporting Fig. S5, we confirm that this gate current $I$ has an Arrhenius relationship with an activation energy equal to 1.2 eV. As we will show in the next section, the channel conductance is controlled by the integral of the gate current $I_G$, or the gate charge $Q$. As a result, if the ionic conductivity and resulting gate current decreases by a factor of 10, then we expect it will take 10 times longer to yield a certain gate charge $Q$, and the retention time will be 10 times higher. Because the ionic resistivity of YSZ has an activation energy of 1.1-1.2 eV (Supporting Fig. S5), we expect that the gate current at 85°C will be about 6,000-13,000 times lower than that at 200°C. This yields an expected retention time from 24 hours at 200°C to 15-35 years at 85°C. We note that this is expected to be a lower bound, as this only represents an extrapolation for our 24-hr experiment at 200°C.

## 2.3 Investigation of Electrochemical Currents

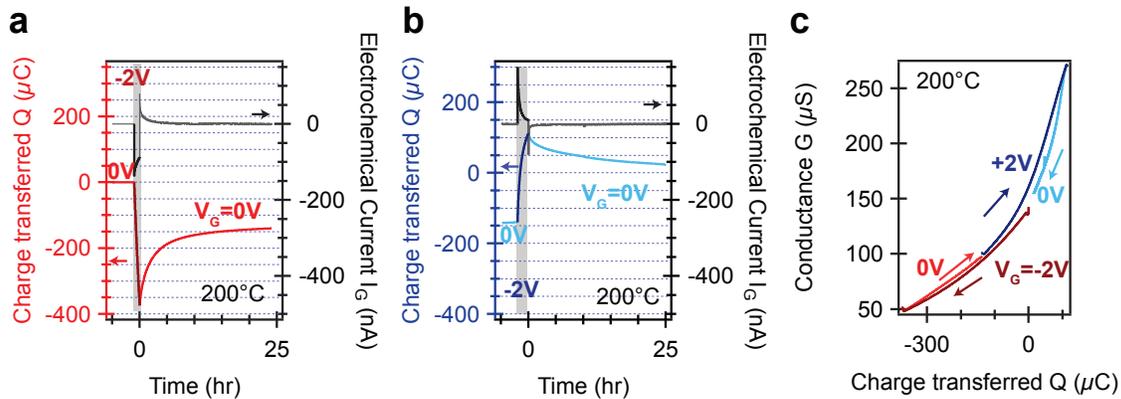

**Fig. 3 The gate electrochemical charge ($Q$) and currents ($I_G$) during switching and retention.** (a) The gate current is negative when -2V is applied, and rebounds during retention. However, the net charge transferred $Q$ converges to -140μC even after 24 hours, suggesting a nonvolatile change in the ion concentration of the channel. (b) A +2V potentiation is applied after the measurements in (a). Once again, the net charge transferred after 24 hours once again appears to converge to a different value, suggesting a nonvolatile change in the channel ion concentration despite the 24 hour retention measurement at short circuit. (c) The electrochemical charge transferred $Q$ directly maps to the channel conductance $G$, consistent with why both these values are nonvolatile.

To better understand the mechanism for nonvolatile ECRAM, we monitor the gate electrochemical current during the switching and retention process. In ECRAM cells containing an electrolyte that conducts ions but blocks electrons, thus ionic current through the electrolyte must equal the electronic current passed between gate and channel (via the external programming circuit) $I_G$ due to charge balance. This electrochemical current $I_G$ is proportional to the time derivative of $X_1$ ($I_G \sim \frac{dX_1}{dt}$), defined as the concentration of oxygen vacancy ions in the channel. Similarly, the electrochemical charge $Q$ ($Q(t) = \int I_G(t)dt$) describes the net change in the concentration of ions ($\Delta X_1$) in the channel over time.

In Fig. 3a, we monitor the gate current $I_G$ and charge $Q$ during a 1 hour depression (-2V) followed by a 24 hour retention (0V), during the same experiment as in Fig. 2b. During depression, $I_G<0$ implies that oxygen vacancies move out of the channel, decreasing the channel conductance (Fig. 2b). However, upon applying 0V, the electrochemical current $I_G$ reverses ($I_G>0$), suggesting that the oxygen vacancies now enter the channel, and are responsible for the transient increase in the channel conductance (Fig. 2b). This electrochemical current rapidly decreases and approaches 0. However, even after 24 hours, $Q$ does not revert to 0. Since ($Q \sim \Delta X_1$) < 0, the oxygen vacancy concentration at the end of the 24 hours is lower than the oxygen vacancy concentration before the depression, a result consistent with the nonvolatile decrease in the channel conductance $G$ (Fig. 2b). Furthermore, the electrochemical current $I_G$ has fallen to such a low value that it is unlikely that $Q$ will reach 0. This strongly suggests that the cell has reached a new, permanent change in the ion concentration ($X_1$), just as it reached a different value for $G$ in Fig. 2b.

In Fig. 3b, we show the same behavior during a 2 hour potentiation at $V_G$=+2V followed by a 24 hour retention at 0V, taken immediately after the result in Fig. 3b. The channel conductance during the measurement was plotted in Fig. 2b. We start $Q$ at -140 µC, continuing from the end of the depression-hold in Fig. 3a. In this case, +2V yields a positive current while the following 0V retention yields a negative current. As in the case for the depression in Fig. 3b, the total integrated charge $Q$ during the hold is much lower than that of the potentiation, even as the current is essentially 0. This result is again consistent with non-volatility observed in Fig. 2b after potentiation. In Fig. 3c, we confirm that the conductance of the channel $G$ is effectively only a function of the integrated gate electrochemical charge $Q$, and nearly independent of other variables. Therefore, changes in the charge $Q$ is directly linked to changes in the conductance $G$.

Summarizing our results in this section, we analyze the electrochemical current on the gate to show that holding the cell at 0V for 24 hours fails to recover the $Q$ that was applied during the previous potentiation and depression. This provides additional evidence that the change in the channel conductance shown in Fig. 2 results from a non-volatile change in the oxygen vacancy concentration $X_1$ of the channel.

## 2.4 Phase separation as the origin of nonvolatility in ECRAM

Having shown a nonvolatile change in the channel conductance $G$ and the ion concentration $X_1$ under short circuit, we seek to understand how this nonvolatile ECRAM differs from previously-demonstrated ECRAM, which are all volatile in comparison. Most volatile ECRAM requires an electronic switch to show any retention beyond a few seconds. In this section, we propose one possible mechanism for how the nonvolatile ECRAM in this work differ from past volatile ECRAM.In Fig. 4a,b, we schematically illustrate ECRAM cells in two states. Fig. 4a shows a state where the channel ion concentration $X_1$ equals the gate ion concentration $X_2$. Fig. 4b shows a state where $X_1 \neq X_2$. The underlying mechanism for all ECRAM devices is that, by applying an electrochemical voltage pulse, we can switch the cell from state 1 to state 2, as well as to a number of other analog resistance states.

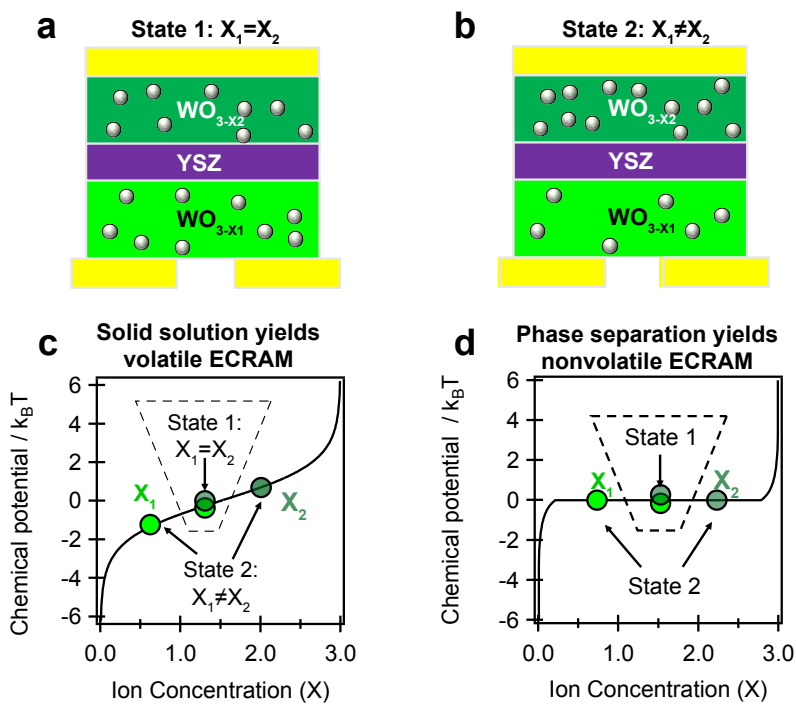

**Fig. 4 Phase separation as a proposed mechanism for the nonvolatile changes in channel ion concentration and electronic conductance.** (a-b) Two states are schematically illustrated indicating the state of the ECRAM cell. $X_1$ and $X_2$ represent the concentration of ions in the channel and gate, respectively. (a) State 1 occurs when $X_1=X_2$ and represents the initial state of the cell that uses the same material for the gate and the channel. (b) State 2 is the general case where $X_1 \neq X_2$, and arises after switching. (c) If the guest ion chemical potential increases monotonically with the ion concentration, then only state 1 is in equilibrium, and state 2 will eventually revert to state 1 over time. This yields volatile ECRAM. (d) If there exists a plateau in the chemical potential, then both state 1 and 2 are stable, implying a nonvolatile change in $X_2$. This nonvolatility arises when the channel material separates into two or more phases. We note that, from a thermodynamic perspective, that there ultimately is only one equilibrium configuration even in two-phase systems due to interfacial energy; however, the driving force to equilibrate interfaces is very small, and will take much longer to equilibrate interfaces than to equilibrate compositional differences in a solid solution.

**Volatile ECRAM based on solid solution**: In previously demonstrated types of ECRAM, it is believed that there is only one equilibrium state. This usually occurs when $X_1=X_2$ (Fig. 4a). When the gate and channel are shorted, the ECRAM cell will revert to this equilibrium state. We propose that ECRAM cells that require a switch should be considered volatile ECRAM because, under equilibrium, they will all revert back to a single resistance state in Fig. 4a. One way to rationalize the relaxation of state 2 to state 1 under short circuit is to consider this a diffusion dominated process where ions migrate from high to low concentrations. Equilibrium is only achieved when $X_1=X_2$, yielding a single equilibrium analog state, and is consistent with drift-diffusion modeling[43].

A more generalizable framework is to recognize that the mass flux follows the spatial gradient of the ion's chemical potential $\mu$, rather than the gradient of the ion concentration $X$. Equilibrium is attained when the ion chemical potential in the channel $\mu(X_1)$ equals that of the gate $\mu(X_2)$. When the chemical potential $\mu(X)$ increases monotonically with X (Fig. 4c), which is the case for when the ion forms a solid solution with the host, this equilibrium condition $\mu(X_1) = \mu(X_2)$ is only attained when $X_1=X_2$. This results in volatile ECRAM with only a single equilibrium state. We note that the exact shape of $\mu(X)$ does not necessarily need to equal that Fig. 4c; rather, it only needs to monotonically increase (or decrease) with the ion concentration X. For example, organic ECRAM operate like electrochemical capacitors[54] where $Q = CV$, but the monotonic relationship between voltage and charge would result in volatile behavior.

**Nonvolatile ECRAM based on phase separation**: As discussed in the previous paragraph, the thermodynamic equilibrium condition is that $\mu(X_1) = \mu(X_2)$. If this condition can be achieved when $X_1 \neq X_2$, such as through the chemical potential profile shown in Fig. 4d, then the switched state B (Fig. 4b) is also in equilibrium and will not revert to state A. In fact, there are likely a very large, possibly continuous, number of analog equilibrium or near-equilibrium states (see caption in Fig. 4d for details). The flat chemical potential profile shown in Fig. 4d is commonly seen in phase-separating materials systems, where the common tangent construction of the Gibbs free energy yield phase separation (Supporting Fig. S6). Such phase-separating systems cannot be modeled using drift-diffusion type equations[39] because ions do not necessarily move from high to low concentrations[55,56]. While phase separation is difficult to probe in amorphous materials like our $WO_{3-X}$ channel, our result nonetheless suggests that phase separation is a strong candidate for the origin of nonvolatility (Fig. 3,4). It is well documented in other electrochemical systems like Li-ion batteries that phase separation is responsible for the simultaneous coexistence of ion-rich and ion-poor regions[57–59]; similarly, in nonvolatile ECRAM, phase separation can yield a gate and a channel with different ion concentrations ($X_1 \neq X_2$) that will not revert to $X_1=X_2$.

Within this framework, we revisit our experimental results that show both a transient volatile response and a nonvolatile response. In Fig. 2b and 3b, the channel conductance $G$ and the gate charge $Q$ decrease during depression but increases during the 24 hour retention. However, this transient increase ($\Delta Q$=+200 μC) is substantially less than the original decrease ($\Delta Q$=-380 μC), and both $Q$ and $G$ ultimately settle to a state different from the initial $Q$ and $G$. This is consistent with state 2 in that Fig. 4b represents a different resistance than state 1 in Fig. 4a; however, in this case, both states are considered equilibrium configurations. We believe that the steady-state, nonvolatile response is a result of phase separation, as $\mu(X_1) = \mu(X_2)$, both at the start and at the end of the measurement. The origins of the long transient response is not clear; some possibilities include the volatile response of an electrochemical double-layer or the nucleation time for phase separation.

Finally, we note that our description in Fig. 4d does not take into account the requirement to minimize interfaces in spinodal decomposition[55,60]. In this respect, there may only be one final, ultimate equilibrium configuration. However, to reach this final state, which may not be state 1 ($X_1=X_2$), it will take much longer than the timescales based on Fickian diffusion, and may not be reachable in experimental timescales. This process may be analogous to coarsening, which slows over time[61]. It is also unclear how these "coarsening" process will affect the measured electronic resistance, making it very difficult to understand the dynamics of reaching this final equilibrium state.

## 2.5 Discussion

Nonvolatile behavior and potentially permanent changes in channel conductance are paradigm changing for ECRAM and nonvolatile memory, especially considering the poor retention of previous reports at scale. The possibility of an ECRAM cell with multiple nonvolatile resistance states under short circuit is compelling for robust inference engines that operate throughout the lifetime of a product. The retention of such cells is orders of magnitude higher than past ECRAM research, and is comparable to that of $HfO_X$-based ReRAM at 200°C (Fig. 2c). Moreover, this work shows that trying to find phase separating materials is the key to achieving ECRAM cells with excellent retention times. Our long retention times show that applications for ECRAM extend beyond on-line training, as suggested in past works[23,28,35]. Inference is not only more broadly applicable, but also one where requirements like switching time, temperature, and endurance are less stringent than for on-line training.

While the nonvolatile $WO_X$-ECRAM is highly promising, there are several materials challenges that must be resolved. The first need is to replace a single-crystal YSZ electrolyte used in this work with a thin-film electrolyte that can be sputter deposited on Si. We have included preliminary results of a $WO_{3-X}$/YSZ/$WO_{3-X}$ ECRAM cell with a thin-film electrolyte with evidence of nonvolatile state retention which is the subject of future work (Supporting Fig. S7); however, additional work is required to achieve high reliability and reproducibility of this device. Creating a thin-film device is not only more practical for CMOS integration, but also substantially reduces the switching and settling times. Because the changes in the conductance states can be permanent due to phase separation (Fig. 4d), we anticipate thin-film devices will also have excellent state retention, even if the ionic currents are higher due to thinner electrolytes. Other tasks include identifying new phase-separating materials with faster switching and shorter settling times. One such material is the well-studied $SrCoO_{2.5}$, which undergoes a phase transition from the Brownmillerite $SrCoO_{2.5}$ to the perovskite $SrCoO_3$ phase[62]. We anticipate that further developments along these fronts may yield permanently nonvolatile ECRAM that can meet all requirements for analog inference.

### 3. <u>Conclusions</u>

In this work, we develop nonvolatile ECRAM using $WO_{3-X}$ where the change in the ion concentration and resistance change in the channel is permanent under short circuit. This likely results from phase separation, whereby the chemical potential of oxygen in the gate and channel are equal, even though their concentrations differ. Our results demonstrate several orders of magnitude increase in the retention time compared to past work, and matches that of state-of-the-art filament-based resistive memory technologies. This work opens a new paradigm for non-volatile ECRAM, as well as highlights a design principle for identifying nonvolatile ECRAM materials.

## 4. Experimental Section

**4.1 Fabrication of $WO_{3-X}$ ECRAM cells**: We purchase 300-μm-thick, 1 cm square, single-crystal yttrium-stabilized zirconia (8 mol% $Y_2O_3$ in $ZrO_2$) substrates from MTIXTL and were used without further cleaning or treatment. We sputter 50 nm of $WO_{3-X}$ on the gate side through a 8 mm square shadow mask using an AJA Orion-8 sputter system. The target was a 3-inch W target (99.95% purity) from Plasmaterials (Livermore, CA, USA). The target power was 100 W. To make the sub-oxide, the sputter gas was a mixture of $Ar:O_2$ at a ratio of 85:15, controlled by mass flow controllers. The sputter gas pressure was 3 mtorr. The substrate to target distance was about 15 cm.

The 18-nm-thick gate Pt current collector was sputtered on top of the $WO_{3-X}$ layer using the same shadow mask without breaking vacuum. The target was a 2-inch Pt target; the sputter gas was 3 mtorr of pure argon. After growing both the suboxide and metal layers, we flip the substrate and grow the same $WO_{3-X}$ on the channel side with the same shadow mask. Finally, the two Pt current collectors on the channel with a different shadow mask (Supporting Fig. S2). This mask has two square regions separated by 0.5 mm. All shadow masks were made from a 0.13 mm thick stainless steel sheet, and cut using a $CO_2$ laser by Stencils Unlimited (Tualatin, OR, USA). All sputtering was done at room temperature; no additional annealing was conducted.

Finally, to protect the cell from environmental oxidation, we use atomic layer deposition to grow 35 nm of $HfO_X$ on the channel side of the ECRAM device. The $HfO_X$ layer was deposited using a Veeco Fiji ALD plasma/thermal assisted atomic layer deposition at the Lurie Nanofabrication Facility. A thermal deposition was done at 200 °C, using the precursor TDMAH and Argon gas, and ran for 297 cycles, at a rate of ~1 Å/cycle and 24 seconds per cycle. The film thickness was measured to be ~35 nm (Fig. 1a). While the $HfO_X$ passivation layer reduced environmental oxidation, it did not fully eliminate oxidation at these elevated temperatures, as seen in the long retention measurements Fig. 2.

**4.2 Device Measurements**. After fabrication, the cell was placed in a custom-built, six-probe Nextron MPS-Ceramic Heater CHH750, environmentally controlled probe station. The two electrodes on the channel directly contacted the Rh probes. The gate current collector rested on a Si substrate with 50 nm of Pt sputtered, and the Rh probe contacts the Pt. The environment was further controlled by flowing 5N ultra high purity Ar at a flow rate of 87 standard cubic centimeters per minute (sccm) controlled by an Omega mass flow controller. A Swagelok check valve was placed to pressurize the chamber to ~ 50 torr (1 pound per square inch) above ambient, which prevents backflow of air into the chamber. A portable Zirox ZR5 oxygen sensor showed that the oxygen concentration in the Ar flow gas is about 3 parts per million, and likely caused the oxidation shown in Fig. 2. We note that flowing Ar was much more effectively at forming a thermal contact between the heater and the chip, compared to pulling a vacuum.

Device measurements were conducted using a Bio-logic SP300 bipotentiostat. Channel was used to apply the gate voltage, while channel 2 was used to measure the channel conductance. The channel conductance was measured by alternating between +10 mV and -10 mV for 30 or 60 seconds each, then taking the average absolute current averaged over 60 or 120 seconds.

**4.3 TEM Measurements**: STEM measurements at the University of Michigan were taken using a Thermo Fisher Talos F200X G2, a 200 kV FEG scanning transmission electron microscope operated in STEM mode. The Velox software was used for STEM images and EDS data acquisitions. The TEM specimen was prepared using a Thermo-Fisher Helios 650 Xe Plasma FIB. The final beam condition was set at 12keV 10 pA for the liftout polishing.

TEM Sample preparation at Sandia National Labs was performed with a ThermoFisher Helios Nanolab 660 Dualbeam instrument. The sample was extracted and thinned with 30 kV $Ga^+$ ions and then final-polished at 5 kV. SNL STEM analysis was performed with a monochromated, aberration corrected ThermoFisher Titan Themis Z operated at 300 kV with a SuperX EDS detector and Gatan Quantum 969 GIF.

**Funding Acknowledgements:**

This work at the University of Michigan was supported by the National Science Foundation under Grant no. ECCS-2106225, startup funding from the University of Michigan College of Engineering, and sub-contracts from Sandia National Laboratories under the Diversity Initiative and Campus Executive Programs. The work at Sandia National Labs was supported the Laboratory-Directed Research and Development (LDRD) program. Sandia National Laboratories is a multi-mission laboratory managed and operated by National Technology and Engineering Solutions of Sandia, LLC, a wholly owned subsidiary of Honeywell International Inc., for the U.S. Department of Energy's National Nuclear Security Administration under contract DE-NA-0003525. A. A. Talin was partly supported by the DOE Office of Science Research Program for Microelectronics Codesign (sponsored by ASCR, BES, HEP, NP, and FES) through the Abisko Project, PM Robinson Pino (ASCR). This paper describes objective technical results and analysis. Any subjective views or opinions that might be expressed in the paper do not necessarily represent the views of the U.S. Department of Energy or the United States Government. The authors acknowledge the University of Michigan College of Engineering for financial support and the Michigan Center for Materials Characterization for use of the instruments and staff assistance. Part of this work was conducted in part of the University of Michigan Lurie Nanofabrication Facility.

**Author Contributions:**
D. K., V. W., L. C., and J. L. fabricated the devices and conducted the measurements. K. S. and J. D. S. conducted TEM measurements. All authors contributed to writing and revising the manuscript.

**Conflicts of Interest:**

The authors declare no financial conflicts of interest.

**Data Availability:**

All the data for the main text are available on the Materials Common Database at https://doi.org/10.13011/m3-tz6h-ht18. Data for the Supporting Information will be available upon reasonable request to the corresponding author.

[1] A. Mehonic, A. J. Kenyon, *Nature* **2022**, *604*, 255.
[2] D. Ielmini, H.-S. P. Wong, *Nat Electron* **2018**, *1*, 333.
[3] Q. Xia, J. J. Yang, *Nat. Mater.* **2019**, *18*, 309.
[4] W. Wan, R. Kubendran, C. Schaefer, S. B. Eryilmaz, W. Zhang, D. Wu, S. Deiss, P. Raina, H. Qian, B. Gao, S. Joshi, H. Wu, H.-S. P. Wong, G. Cauwenberghs, *Nature* **2022**, *608*, 504.
[5] M. J. Marinella, S. Agarwal, A. Hsia, I. Richter, R. Jacobs-Gedrim, J. Niroula, S. J. Plimpton, E. Ipek, C. D. James, *IEEE Journal on Emerging and Selected Topics in Circuits and Systems* **2018**, *8*, 86.
[6] J. Hasler, *Proc. IEEE* **2020**, *108*, 1283.
[7] S. Ambrogio, P. Narayanan, H. Tsai, C. Mackin, K. Spoon, A. Chen, A. Fasoli, A. Friz, G. W. Burr, in *2020 International Symposium on VLSI Technology, Systems and Applications (VLSI-TSA)*, **2020**, pp. 119–120.
[8] S. Ambrogio, P. Narayanan, H. Tsai, R. M. Shelby, I. Boybat, C. di Nolfo, S. Sidler, M. Giordano, M. Bodini, N. C. P. Farinha, B. Killeen, C. Cheng, Y. Jaoudi, G. W. Burr, *Nature* **2018**, *558*, 60.
[9] C. Mackin, M. J. Rasch, A. Chen, J. Timcheck, R. L. Bruce, N. Li, P. Narayanan, S. Ambrogio, M. Le Gallo, S. R. Nandakumar, A. Fasoli, J. Luquin, A. Friz, A. Sebastian, H. Tsai, G. W. Burr, *Nat Commun* **2022**, *13*, 3765.
[10] A. I. Khan, A. Keshavarzi, S. Datta, *Nat Electron* **2020**, *3*, 588.
[11] P. Wang, S. Yu, *MRS Communications* **2020**, *10*, 538.
[12] Z. Wang, H. Wu, G. W. Burr, C. S. Hwang, K. L. Wang, Q. Xia, J. J. Yang, *Nat Rev Mater* **2020**, *5*, 173.
[13] P. Yao, H. Wu, B. Gao, J. Tang, Q. Zhang, W. Zhang, J. J. Yang, H. Qian, *Nature* **2020**, *577*, 641.
[14] M. Hu, C. E. Graves, C. Li, Y. Li, N. Ge, E. Montgomery, N. Davila, H. Jiang, R. S. Williams, J. J. Yang, Q. Xia, J. P. Strachan, *Advanced Materials* **2018**, *30*, 1705914.
[15] T. Stecconi, R. Guido, L. Berchialla, A. La Porta, J. Weiss, Y. Popoff, M. Halter, M. Sousa, F. Horst, D. Dávila, U. Drechsler, R. Dittmann, B. J. Offrein, V. Bragaglia, *Advanced Electronic Materials* **n.d.**, 2200448.
[16] R. Dittmann, S. Menzel, R. Waser, *Advances in Physics* **2022**, *70*, 155.
[17] R. Waser, R. Dittmann, G. Staikov, K. Szot, *Adv. Mater.* **2009**, *21*, 2632.
[18] S. Liu, T. P. Xiao, C. Cui, J. A. C. Incorvia, C. H. Bennett, M. J. Marinella, *Appl. Phys. Lett.* **2021**, *118*, 202405.
[19] J. Shi, S. D. Ha, Y. Zhou, F. Schoofs, S. Ramanathan, *Nat Commun* **2013**, *4*, 2676.
[20] J. D. Greenlee, C. F. Petersburg, W. G. Daly, F. M. Alamgir, W. Alan Doolittle, *Appl. Phys. Lett.* **2013**, *102*, 213502.
[21] E. J. Fuller, F. E. Gabaly, F. Léonard, S. Agarwal, S. J. Plimpton, R. B. Jacobs-Gedrim, C. D. James, M. J. Marinella, A. A. Talin, *Advanced Materials* **2017**, *29*, 1604310.
[22] Y. van de Burgt, E. Lubberman, E. J. Fuller, S. T. Keene, G. C. Faria, S. Agarwal, M. J. Marinella, A. Alec Talin, A. Salleo, *Nature Mater* **2017**, *16*, 414.
[23] E. J. Fuller, S. T. Keene, A. Melianas, Z. Wang, S. Agarwal, Y. Li, Y. Tuchman, C. D. James, M. J. Marinella, J. J. Yang, A. Salleo, A. A. Talin, *Science* **2019**, 570.

[24] J. Tang, D. Bishop, S. Kim, M. Copel, T. Gokmen, T. Todorov, S. Shin, K.-T. Lee, P. Solomon, K. Chan, W. Haensch, J. Rozen, in *2018 IEEE International Electron Devices Meeting (IEDM)*, **2018**, p. 13.1.1-13.1.4.
[25] Y. Li, E. J. Fuller, J. D. Sugar, S. Yoo, D. S. Ashby, C. H. Bennett, R. D. Horton, M. S. Bartsch, M. J. Marinella, W. D. Lu, A. A. Talin, *Advanced Materials* **2020**, *32*, 2003984.
[26] Y. Li, E. J. Fuller, S. Asapu, S. Agarwal, T. Kurita, J. J. Yang, A. A. Talin, *ACS Appl. Mater. Interfaces* **2019**, *11*, 38982.
[27] S. Kim, T. Todorov, M. Onen, T. Gokmen, D. Bishop, P. Solomon, K.-T. Lee, M. Copel, D. B. Farmer, J. A. Ott, T. Ando, H. Miyazoe, V. Narayanan, J. Rozen, in *2019 IEEE International Electron Devices Meeting (IEDM)*, **2019**, p. 35.7.1-35.7.4.
[28] M. Onen, N. Emond, B. Wang, D. Zhang, F. M. Ross, J. Li, B. Yildiz, J. A. Del Alamo, *Science* **2022**, *377*, 539.
[29] X. Yao, K. Klyukin, W. Lu, M. Onen, S. Ryu, D. Kim, N. Emond, I. Waluyo, A. Hunt, J. A. del Alamo, J. Li, B. Yildiz, *Nat Commun* **2020**, *11*, 3134.
[30] J. Lee, R. D. Nikam, M. Kwak, H. Hwang, *ACS Appl. Mater. Interfaces* **2022**, *14*, 13450.
[31] J. Lee, R. D. Nikam, M. Kwak, H. Kwak, S. Kim, H. Hwang, *Advanced Electronic Materials* **2021**, *7*, 2100219.
[32] R. D. Nikam, M. Kwak, H. Hwang, *Advanced Electronic Materials* **2021**, *7*, 2100142.
[33] R. D. Nikam, J. Lee, W. Choi, D. Kim, H. Hwang, *ACS Nano* **2022**, acsnano.2c02913.
[34] A. Melianas, M.-A. Kang, A. VahidMohammadi, T. J. Quill, W. Tian, Y. Gogotsi, A. Salleo, M. M. Hamedi, *Advanced Functional Materials* **2022**, *32*, 2109970.
[35] Y. Li, T. P. Xiao, C. H. Bennett, E. Isele, A. Melianas, H. Tao, M. J. Marinella, A. Salleo, E. J. Fuller, A. A. Talin, *Front Neurosci* **2021**, *15*, 636127.
[36] M. T. Sharbati, Y. Du, J. Torres, N. D. Ardolino, M. Yun, F. Xiong, *Advanced Materials* **2018**, *30*, 1802353.
[37] D. Kireev, S. Liu, H. Jin, T. Patrick Xiao, C. H. Bennett, D. Akinwande, J. A. C. Incorvia, *Nat Commun* **2022**, *13*, 4386.
[38] Y. Tuchman, T. J. Quill, G. LeCroy, A. Salleo, *Advanced Electronic Materials* **n.d.**, *n/a*, 2100426.
[39] E. J. Fuller, Y. Li, C. Bennet, S. T. Keene, A. Melianas, S. Agarwal, M. J. Marinella, A. Salleo, A. A. Talin, *IBM Journal of Research and Development* **2019**, *63*, 9:1.
[40] S. T. Keene, A. Melianas, Y. van de Burgt, A. Salleo, *Advanced Electronic Materials* **2019**, *5*, 1800686.
[41] Y. Li, J. Lu, D. Shang, Q. Liu, S. Wu, Z. Wu, X. Zhang, J. Yang, Z. Wang, H. Lv, M. Liu, *Advanced Materials* **2020**, *32*, 2003018.
[42] Q. Wan, M. Rasetto, M. T. Sharbati, J. R. Erickson, S. R. Velagala, M. T. Reilly, Y. Li, R. Benosman, F. Xiong, *Advanced Intelligent Systems* **2021**, *3*, 2100021.
[43] M. Baldo, D. Ielmini, in *2021 IEEE International Memory Workshop (IMW)*, **2021**, pp. 1–4.
[44] Y. Jeong, H. Lee, D. G. Ryu, S. H. Cho, G. Lee, S. Kim, S. Kim, Y. S. Lee, *Advanced Electronic Materials* **2021**, *7*, 2100185.
[45] S. Kazemzadeh, L. Dodsworth, I. F. Pereira, Y. van de Burgt, *Advanced Electronic Materials* **n.d.**, *n/a*, 2200427.
[46] M. Onen, N. Emond, J. Li, B. Yildiz, J. A. del Alamo, *Nano Lett.* **2021**, *21*, 6111.
[47] A. Melianas, T. J. Quill, G. LeCroy, Y. Tuchman, H. v. Loo, S. T. Keene, A. Giovannitti, H. R. Lee, I. P. Maria, I. McCulloch, A. Salleo, *Sci. Adv.* **2020**, *6*, eabb2958.

[48] T. Gokmen, Y. Vlasov, *Front. Neurosci.* **2016**, *10*, DOI 10.3389/fnins.2016.00333.
[49] S. T. Keene, A. Melianas, E. J. Fuller, Y. van de Burgt, A. A. Talin, A. Salleo, *J. Phys. D: Appl. Phys.* **2018**, *51*, 224002.
[50] B. Traoré, P. Blaise, E. Vianello, H. Grampeix, S. Jeannot, L. Perniola, B. De Salvo, Y. Nishi, *IEEE Transactions on Electron Devices* **2015**, *62*, 4029.
[51] Y. Y. Chen, R. Degraeve, S. Clima, B. Govoreanu, L. Goux, A. Fantini, G. S. Kar, G. Pourtois, G. Groeseneken, D. J. Wouters, M. Jurczak, in *2012 International Electron Devices Meeting*, **2012**, p. 20.3.1-20.3.4.
[52] Z. Wei, T. Takagi, Y. Kanzawa, Y. Katoh, T. Ninomiya, K. Kawai, S. Muraoka, S. Mitani, K. Katayama, S. Fujii, R. Miyanaga, Y. Kawashima, T. Mikawa, K. Shimakawa, K. Aono, in *2012 4th IEEE International Memory Workshop*, **2012**.
[53] Y. Y. Chen, L. Goux, S. Clima, B. Govoreanu, R. Degraeve, G. S. Kar, A. Fantini, G. Groeseneken, D. J. Wouters, M. Jurczak, *IEEE Transactions on Electron Devices* **2013**, *60*, 1114.
[54] C. M. Proctor, J. Rivnay, G. G. Malliaras, *Journal of Polymer Science Part B: Polymer Physics* **2016**, *54*, 1433.
[55] M. Z. Bazant, *Acc. Chem. Res.* **2013**, *46*, 1144.
[56] H. Tian, M. Z. Bazant, *Nano Lett.* **2022**, *22*, 5866.
[57] Y. Li, W. C. Chueh, *Annual Review of Materials Research* **2018**, *48*, 137.
[58] Y. Li, H. Chen, K. Lim, H. D. Deng, J. Lim, D. Fraggedakis, P. M. Attia, S. C. Lee, N. Jin, J. Moškon, Z. Guan, W. E. Gent, J. Hong, Y.-S. Yu, M. Gaberšček, M. S. Islam, M. Z. Bazant, W. C. Chueh, *Nature Mater* **2018**, *17*, 915.
[59] J. Lim, Y. Li, D. H. Alsem, H. So, S. C. Lee, P. Bai, D. A. Cogswell, X. Liu, N. Jin, Y. Yu, N. J. Salmon, D. A. Shapiro, M. Z. Bazant, T. Tyliszczak, W. C. Chueh, *Science* **2016**, *353*, 566.
[60] J. W. Cahn, J. E. Hilliard, **n.d.**, 11.
[61] B. König, O. J. J. Ronsin, J. Harting, *Phys. Chem. Chem. Phys.* **2021**, *23*, 24823.
[62] Q. Lu, B. Yildiz, *Nano Lett.* **2016**, *16*, 1186.

Supporting information for

**Nonvolatile Electrochemical Random-Access Memory Under Short Circuit**

Diana Kim[1,2], Virgil Watkins[1], Laszlo Cline[1], Jingxian Li[1], Kai Sun[1],
Joshua D. Sugar[3], Elliot J. Fuller[3], A. Alec Talin[3], Yiyang Li[1#]

[1]Materials Science and Engineering, University of Michigan, Ann Arbor, MI, USA

[2]Macromolecular Science and Engineering, University of Michigan, Ann Arbor, MI, USA

[3] Sandia National Laboratories, Livermore, CA, USA

#Corresponding author: yiyangli@umich.edu

Supporting Figures S1-S7

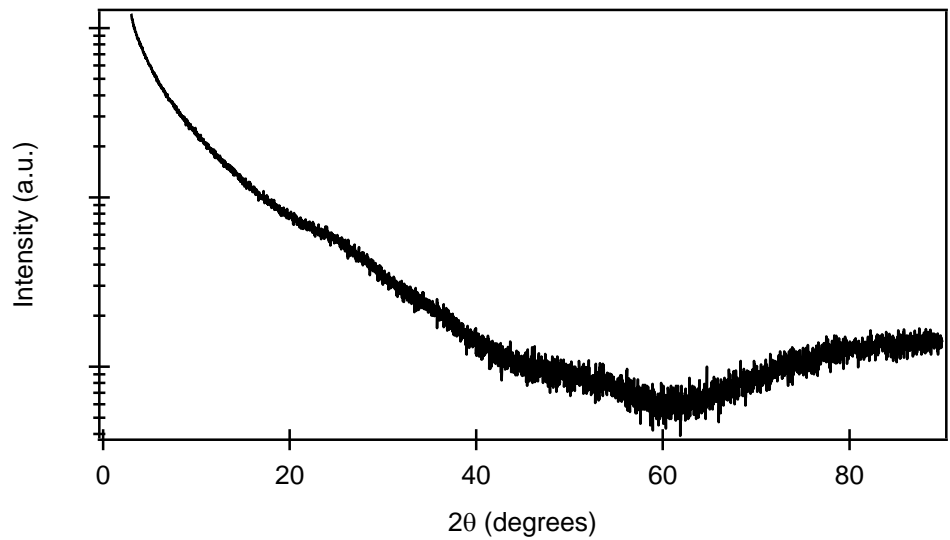

**Supporting Fig. S1**: X-ray diffraction profile of $WO_{3-X}$ confirms that the channel material is amorphous.

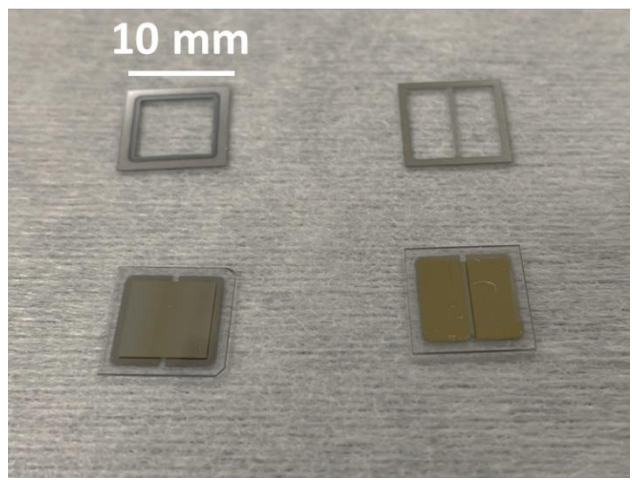

**Supporting Fig. S2**: Images of the shadow masks and the ECRAM cells used in this study.

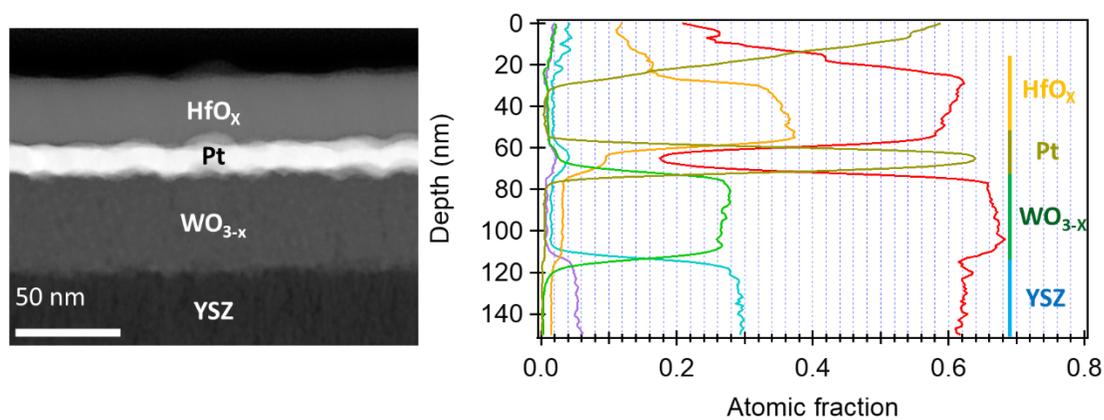

**Supporting Fig. S3**: STEM and energy dispersive spectroscopy linescan of an ECRAM device shows that the W:O ratio in the $WO_{3-X}$ film is approximately 1:2.5 (X~0.5).

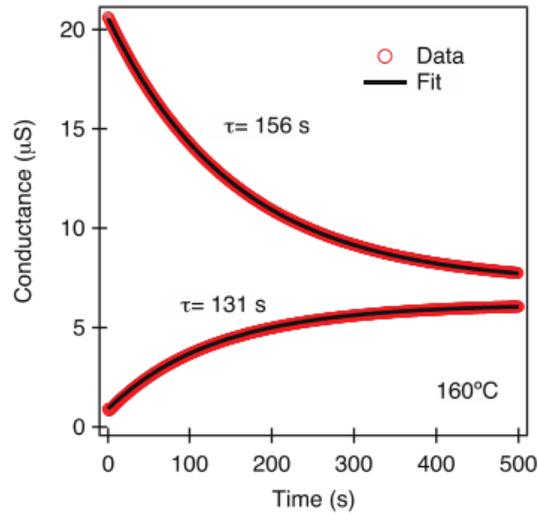

**Supporting Fig. S4**: Retention behavior of TiO$_2$-based ECRAM shows rapid convergence to a single value from the high and low resistance state. Data and image from ref. [23].

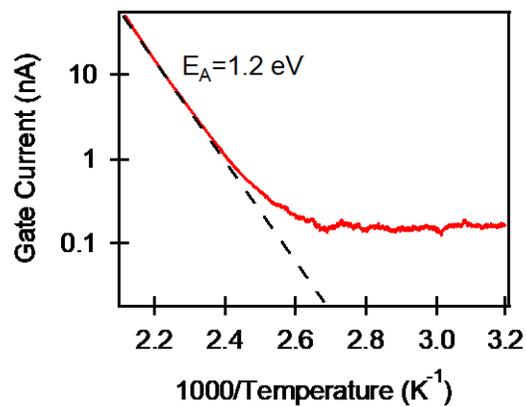

**Supporting Fig. S5:** The gate electrochemical current upon the application of +2V as a function of temperature shows an Arrhenius relationship with an activation energy of 1.2 eV. This result is consistent with the measured YSZ activation energy of 1.1 eV in previous work. [23]. We note that the currents at lower temperatures is not accurate due to the limitations of the instrument.

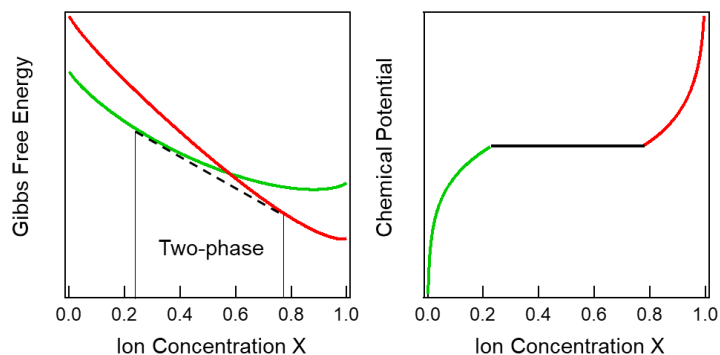

**Supporting Fig. S6:** Phase separation when the common tangent of the Gibbs Free Energy has a lower energy than that of one constituent phase. This common tangent means that the chemical potential, or the slope of the Gibbs Free Energy, is invariant with the ion concentration X.

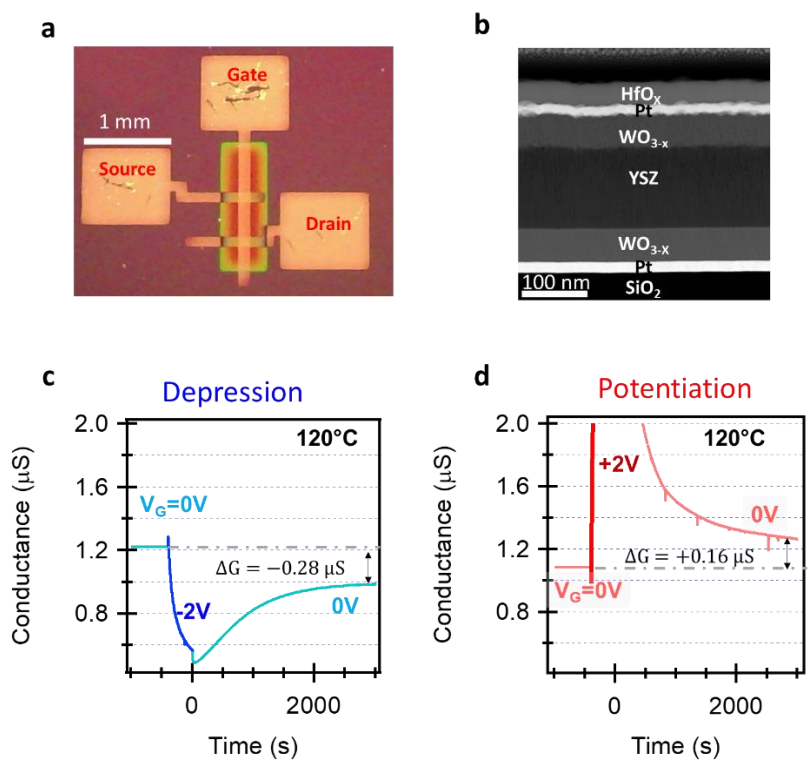

**Supporting Fig. S7:** Preliminary demonstration of thin-film ECRAM cell using the same $WO_{3-X}$ gate and channel. (a) Optical micrograph of the thin-film ECRAM. The rectangular region in the middle contains the gate, YSZ electrolyte (grown by RF sputtering at 100W for a YSZ target and under 3 mtorr of Ar), and channel stacks. (b) Scanning transmission electron microscopy cross-section of the device. (c,d) Potentiation and depression of the cell show some nonvolatile state retention.